\documentclass[twocolumn,english,prd,showpacs,superscriptaddress,longbibliography]{revtex4-1}
\usepackage{amsfonts}
\usepackage{amsmath}
\usepackage{amssymb}
\usepackage[colorlinks, citecolor=blue,linkcolor=red]{hyperref}
\usepackage{color}
\usepackage{textcomp}
\usepackage[rightcaption]{sidecap}
\usepackage{subfigure}
\usepackage[utf8]{inputenc}
\usepackage[T1]{fontenc}
\usepackage{graphicx}
\usepackage{lipsum}

\begin{document}
	\title{Horizon-Scale Corrections to Black Hole Geometry from a Conserved Quantum Atmosphere}
	\author{Kashif Ammar Yasir}
	\email{kayasir@zjnu.edu.cn}
	\affiliation{Department of Physics, Zhejiang Normal University, Jinhua 321004, China.}
	\affiliation{Zhejiang Institute of Photoelectronics, Jinhua 321004, China.}
	\setlength{\parskip}{0pt}
	\setlength{\belowcaptionskip}{-10pt}
	
\begin{abstract}
Black holes are usually described either as classical geometries on which quantum fields propagate or as objects whose complete quantum description lies beyond semiclassical gravity. Hawking radiation already shows, however, that a black hole is not an exactly isolated system: the horizon carries a thermal scale and, after coarse graining, may be associated with fluctuations, dissipation and environmental correlations. Whether this horizon-scale openness can be encoded directly in the gravitational background, before any stochastic treatment of the metric, remains an open question. Here we answer it by introducing, for the first time, a conserved Hawking-scale anisotropic source representing the coarse-grained near-horizon atmosphere and solving the Einstein equations for a static black hole at fixed ADM mass. The dressed geometry has a renormalized mass function, a self-consistently shifted horizon and a modified surface gravity. Because the same lapse governs the horizon and timelike geodesics, the thermal response is accompanied by a correlated orbital restructuring: the effective potential is renormalized, the ISCO moves inward and the near-horizon residence time increases. The construction adds no Langevin force to a fixed Schwarzschild background; rather, it defines a deterministic dressed mean geometry, providing a controlled background on which stochastic metric fluctuations may subsequently be treated in the Einstein–Langevin framework. This establishes the deterministic mean-field sector as the necessary first step for future stochastic gravity programs addressing horizon-scale openness.
\end{abstract}

	\date{\today}
	\maketitle
\section{Introduction}

Black holes provide one of the most direct arenas in which general relativity, quantum field theory and thermodynamics meet. In classical gravity, a stationary black hole is characterized by its geometry, horizon structure and conserved charges. In quantum field theory on curved spacetime, however, the same object acquires a thermal character: Hawking radiation assigns a temperature to the horizon and shows that the black hole cannot be regarded as a perfectly isolated classical system \cite{Einstein1915,Schwarzschild1916,Kerr1963,Carter1968,Bardeen1973,Hawking1975,Unruh1976,BirrellDavies,Wald1994}. This changes the physical status of the horizon. It is not only a causal boundary of a classical spacetime, but also a thermodynamic interface where geometry, quantum fluctuations and coarse-grained irreversibility are tied together.

This viewpoint has developed along several complementary directions. Semiclassical gravity describes the response of a classical metric to quantum expectation values of matter fields, while stochastic gravity and Einstein--Langevin formulations extend the description beyond the mean field by allowing stress-tensor fluctuations to induce metric fluctuations \cite{CaldeiraLeggett,HuPazZhang,HuVerdaguer}. More generally, open-system methods provide a natural language for coarse graining, dissipation, noise and fluctuation--dissipation relations in quantum systems. Related quantum-Langevin approaches have also been useful in driven and dissipative platforms, including cavity and hybrid systems, where coupling to an environment can reorganize collective dynamics, stability and topology \cite{kashif4,kashif42,Refb,Yasir2026Book,kashif26}. In black-hole physics, the information problem further reinforces the idea that horizons cannot be treated as passive classical surfaces once Hawking radiation, entropy and quantum correlations are included \cite{Page1980,Page1993,HaydenPreskill,Marolf2017,UnruhWald2017,Almheiri2021,BalasubramanianPRL2024,BalasubramanianPRX2024}. At the same time, analogue-horizon experiments and quantum-noise-limited measurements have made fluctuation-driven aspects of horizon physics increasingly operational \cite{Philbin2008,Steinhauer2016,MunozdeNova2019,Kolobov2021,Patrick2021,Jia2024,Cromb2025}.

Despite this progress, there remains a useful intermediate problem between fixed-background semiclassical gravity and a full microscopic theory of quantum spacetime. One may ask whether the thermal scale set by Hawking radiation can be incorporated into an effective gravitational background itself, before solving the complete stochastic dynamics of the metric. Such a construction must be more constrained than an arbitrary deformation of the Schwarzschild solution. The effective source that dresses the metric must satisfy the Einstein equations, respect stress-energy conservation and admit a controlled limit in which the Schwarzschild geometry is recovered. If these conditions are imposed, the resulting mean geometry can provide a tractable setting in which horizon-scale open-system effects are encoded directly in the spacetime background.

In this work we develop such a semiclassical mean-field construction for a static black hole. The coarse-grained near-horizon atmosphere is represented by a Hawking-scale anisotropic source, and the source is required to be conserved by the Bianchi identity. This condition fixes the allowed pressure balance and prevents the dressing from being an unconstrained phenomenological profile. The source renormalizes the Schwarzschild seed geometry through a radial mass function. In a conserved lapse-only branch, the additional redshift factor is absent, while the lapse itself is modified. The dressed horizon is then obtained self-consistently from the zero of the corrected lapse, rather than imposed at the reference Schwarzschild radius. The same geometry determines the effective surface gravity, the geodesic potential, the circular-orbit sequence, the innermost stable circular orbit and the near-horizon capture dynamics.

The aim is not to claim a microscopic quantum-gravity solution or to replace the full Einstein--Langevin problem. The present construction isolates the deterministic mean-field sector of a broader open-system description. The stochastic part, encoded in the noise kernel and, in a full linear-response treatment, the associated dissipation kernel, remains the next level of the theory. In this sense, the work provides a controlled semiclassical bridge between fixed-background horizon thermodynamics and fluctuating quantum spacetime: the geometry is first dressed at the mean-field level, while genuine metric fluctuations are to be restored through the Einstein--Langevin extension.

The paper is organized as follows. Section~II introduces the Hawking-scale effective source, derives the conserved mean-field geometry, and defines the dressed horizon, temperature and geodesic sector. Sections~III--VII present the results: the mean geometric dressing, the horizon and thermal response, the orbital restructuring, the correlation between thermal and orbital diagnostics, and the plunge dynamics. Section~VIII summarizes the physical conclusions. The appendices provide the detailed derivations of the reduced Einstein equations, the conservation condition, ADM normalization, horizon-root structure, effective temperature, circular-orbit and ISCO equations, plunge prescription, and the formal Einstein--Langevin extension.
	
	\section{Hawking-scale mean-field dressing}
	\begin{figure*}[htp]
		\includegraphics[width=14cm]{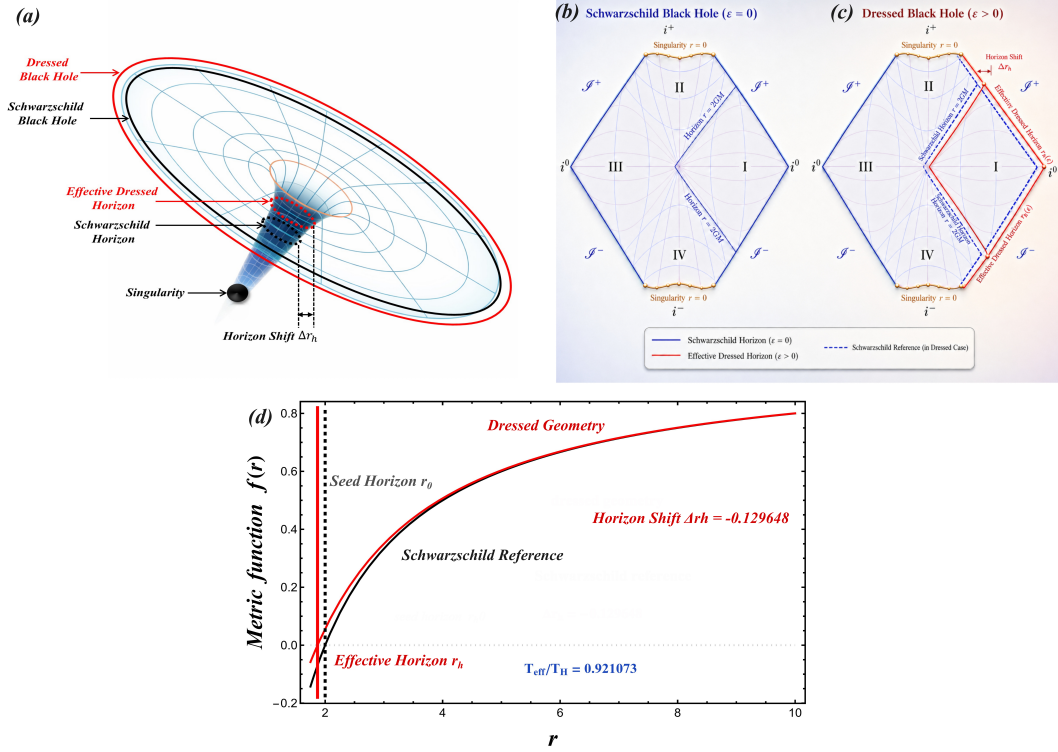}
		\caption{\textbf{Mean-field dressing of the Schwarzschild geometry by a Hawking-scale effective atmosphere.}
			\textbf{(a)} Embedding comparison between the Schwarzschild reference geometry and the dressed mean geometry obtained from the conserved effective source. The red surface denotes the dressed solution and the black surface denotes the Schwarzschild reference with the same ADM mass. The displacement of the zero of the lapse defines the horizon shift \(\Delta r_h=r_h-r_0\), where \(r_0=2M\).
		\textbf{(b)} Schematic horizon-scale diagram of the Schwarzschild reference spacetime.
		\textbf{(c)} Schematic horizon-scale diagram of the dressed mean geometry. The red boundary denotes the effective horizon \(r_h(\varepsilon)\), while the dashed blue line marks the reference Schwarzschild radius \(r_0\). The diagram is intended to display the displacement of the horizon scale; the full stochastic causal structure is not included at this deterministic level.
			\textbf{(d)} Lapse function \(f(r)\) for the dressed and Schwarzschild geometries. For the parameter set shown, the corrected lapse has an inward horizon displacement \(\Delta r_h=-0.129648\), together with an effective temperature ratio \(T_{\mathrm{eff}}/T_H=0.921073\). The figure illustrates that the same conserved Hawking-scale source changes both the horizon location and the thermal response of the mean geometry before stochastic metric fluctuations are restored.}
		\label{fig1}
	\end{figure*}
	
	We now construct the mean geometry used in this work. The guiding assumption is that the black hole is not an exactly isolated classical object once Hawking radiation is admitted. The horizon is a thermodynamic interface, and the degrees of freedom that are coarse grained into this interface can leave a mean imprint on the background geometry before the genuinely stochastic metric fluctuations are restored. We therefore separate the problem into two levels. At the first level, the metric is a deterministic mean field \(\bar g_{\mu\nu}\). At the second level, the same open-system description would be completed by a noise kernel and metric fluctuations. The present section concerns only the first level.
	
	The mean geometry is taken to satisfy
	\begin{equation}
		G_{\mu\nu}[\bar g]=8\pi T_{\mu\nu}^{(\mathrm{env})},
		\label{eq:mean_einstein}
	\end{equation}
	where \(T_{\mu\nu}^{(\mathrm{env})}\) is a coarse-grained effective stress tensor associated with the near-horizon quantum atmosphere. In the present paper this tensor is used as a semiclassical effective source; it is not obtained from a microscopic calculation of a renormalized stress tensor in a specified quantum state. This source is not an arbitrary deformation. Since \(\nabla_\mu G^{\mu\nu}=0\), consistency requires
	\begin{equation}
		\bar\nabla_\mu T^{\mu\nu}_{(\mathrm{env})}=0 .
		\label{eq:main_conservation}
	\end{equation}
	This condition is important physically: it means that the environment may dress the geometry, but it cannot inject or remove energy-momentum in a way that violates the local constraints of general relativity. The construction is therefore an effective conserved backreaction model, not a friction force appended to geodesic motion.
	
	We use the static, spherically symmetric line element
	\begin{equation}
		ds^2=-e^{2\Phi(r)}f(r)\,dt^2+\frac{dr^2}{f(r)}+r^2d\Omega^2,
		\qquad
		f(r)=1-\frac{2m(r)}{r}.
		\label{eq:main_metric}
	\end{equation}
	The two functions have distinct physical meanings. The mass function \(m(r)\) measures how much gravitating energy is enclosed inside radius \(r\), while \(\Phi(r)\) measures any additional redshift not already contained in the lapse \(f(r)\). For an anisotropic effective source
	\begin{equation}
		T^\mu{}_\nu=
		\mathrm{diag}\!\left[-\rho(r),\,p_r(r),\,p_\perp(r),\,p_\perp(r)\right],
		\label{eq:main_stress}
	\end{equation}
	the independent Einstein equations reduce to
	\begin{equation}
		m'(r)=4\pi r^2\rho(r),
		\Phi'(r)=\frac{4\pi r[\rho(r)+p_r(r)]}{f(r)} .
		\label{eq:main_reduced}
	\end{equation}
	Thus the density \(\rho\) changes the radial mass profile, whereas the combination \(\rho+p_r\) decides whether the dressing also produces an extra temporal redshift. This separation is useful because it identifies a controlled lapse-only sector in which the geometry is modified while the normalization of the time coordinate remains Schwarzschild-like at infinity.
	
	The only scale available to a neutral, non-rotating black hole in this problem is the Hawking scale. This profile is used as an exterior effective atmosphere and is not intended as a regular interior solution extending to $r=0$. We therefore take the effective source to be localized near the Schwarzschild reference radius
	\begin{equation}
		r_0=2M,
		\qquad
		T_H=\frac{1}{8\pi M},
		\label{eq:main_hawking_scale}
	\end{equation}
	where \(M\) is the ADM mass. To keep dimensions explicit we write the geometrized energy density as
	\begin{equation}
		\rho(r)=\varepsilon\,\ell_{\rm P}^{\,2}T_H^4
		\left(\frac{r_0}{r}\right)^n,
		\qquad n>3 .
		\label{eq:main_density}
	\end{equation}
	Here \(\ell_{\rm P}\) is the Planck length and \(\varepsilon\) is a dimensionless strength parameter that absorbs numerical factors, greybody weights and the effective number of near-horizon degrees of freedom. In Planck units \(\ell_{\rm P}=1\). The condition \(n>3\) ensures that the atmosphere carries a finite integrated contribution at large radius. The power \(n\) controls how sharply the effective source is concentrated near the horizon.
	
	A simple conserved subclass is obtained by choosing
	\begin{equation}
		p_r(r)=-\rho(r).
		\label{eq:main_pr}
	\end{equation}
	Then \(\rho+p_r=0\), so Eq.~\eqref{eq:main_reduced} gives \(\Phi'(r)=0\). After the usual normalization of the time coordinate at infinity, \(\Phi=0\). The metric therefore reduces to the lapse-only form
	\begin{equation}
		ds^2=-f(r)\,dt^2+\frac{dr^2}{f(r)}+r^2d\Omega^2 .
		\label{eq:main_lapse_metric}
	\end{equation}
	The transverse pressure is then fixed by conservation rather than chosen independently. From \(\bar\nabla_\mu T^{\mu r}=0\), one obtains
	\begin{equation}
		p_\perp(r)=\left(\frac{n}{2}-1\right)\rho(r).
		\label{eq:main_pperp}
	\end{equation}
	This relation is the point at which the effective model becomes a consistent anisotropic source for the Einstein equations. The anisotropy has a clear physical role: the radial vacuum-like pressure \(p_r=-\rho\) prevents extra redshift renormalization, while the transverse pressure balances the radial gradient of the horizon-localized energy density.
	
		\begin{figure*}[htp]
		\includegraphics[width=9cm]{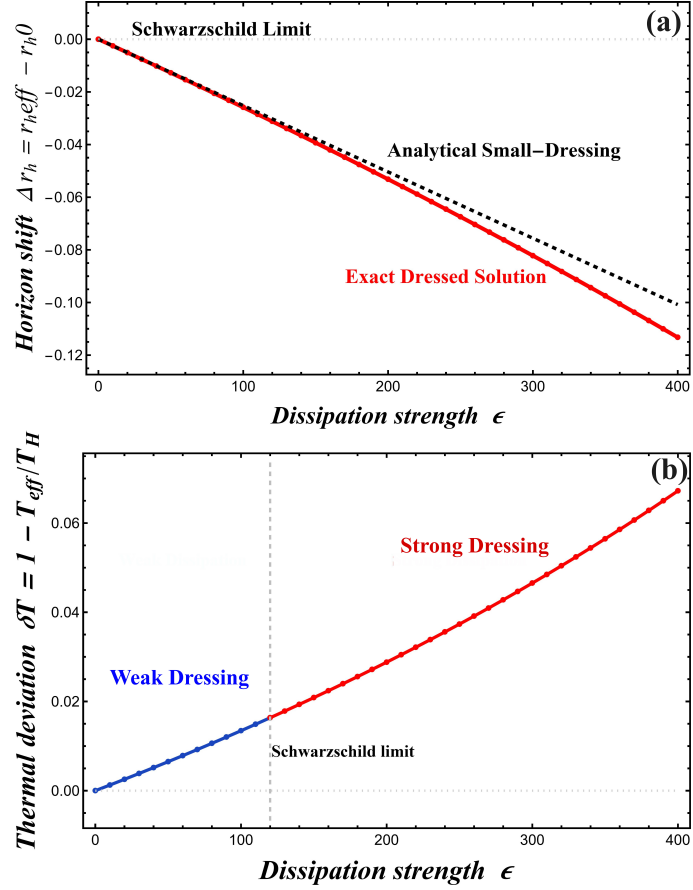}
		\caption{\textbf{Geometric and thermal response of the dressed mean black hole.}
			\textbf{(a)} Horizon shift \(\Delta r_h=r_h-r_0\) as a function of the dissipation/dressing strength \(\varepsilon\). The red curve gives the numerical root of the corrected lapse, while the black dotted curve shows the weak-dressing expansion. Agreement at small \(\varepsilon\) verifies the perturbative limit; the departure at larger \(\varepsilon\) signals nonlinear response of the lapse.
			\textbf{(b)} Thermal deviation \(\delta T=1-T_{\mathrm{eff}}/T_H\) over the same range. The monotonic growth of \(\delta T\) shows that the effective temperature is suppressed as the horizon-localized dressing becomes stronger. The two branches indicate the weak- and strong-dressing regimes used to guide the eye.}
		\label{fig2}
	\end{figure*}
	We fix the ADM mass rather than the mass at the seed horizon. Thus
	\begin{equation}
		m(\infty)=M,
		m(r)=M-\int_r^\infty 4\pi \tilde r^2\rho(\tilde r)\,d\tilde r .
		\label{eq:main_adm}
	\end{equation}
	For the profile in Eq.~\eqref{eq:main_density}, this gives
	\begin{align}
		m(r)&=M-\frac{4\pi\varepsilon\ell_{\rm P}^{\,2}T_H^4 r_0^n}{n-3}\,r^{3-n},\\
		f(r)&=1-\frac{2M}{r}
		+\frac{8\pi\varepsilon\ell_{\rm P}^{\,2}T_H^4 r_0^n}{n-3}\,r^{2-n}.
		\label{eq:main_mass_lapse}
	\end{align}	
	This normalization is essential. If the mass function is instead anchored at \(r_0=2M\), then \(f(r_0)=0\) remains true by construction and no genuine horizon displacement is obtained. With ADM normalization, \(M\) is the total mass seen at infinity, while part of that mass is distributed in the effective atmosphere outside a given radius. The enclosed mass \(m(r)\) is therefore smaller than \(M\) near the horizon, and the outer zero of the lapse can move inward relative to \(r_0\).
	
	The dressed horizon is the largest positive root of
	\begin{equation}
		f(r_h)=0 .
		\label{eq:main_horizon}
	\end{equation}
	For weak dressing, \(r_h=r_0+\delta r_h\), and expansion of Eq.~\eqref{eq:main_mass_lapse} gives
	\begin{equation}
		\delta r_h
		=
		-\frac{8\pi\varepsilon\ell_{\rm P}^{\,2}T_H^4 r_0^3}{n-3}
		+\mathcal{O}(\varepsilon^2).
		\label{eq:main_horizon_shift}
	\end{equation}
	The negative sign should not be interpreted as a loss of ADM mass. It reflects a redistribution of the fixed total mass: some of the gravitating contribution is carried by the atmosphere outside the would-be Schwarzschild radius, so the radius enclosing the horizon mass is reduced.
	
	The same mean geometry fixes the effective temperature. For the lapse-only metric, the surface gravity is \(\kappa=f'(r_h)/2\), and hence
	\begin{equation}
		T_{\rm eff}=\frac{f'(r_h)}{4\pi}
		=
		\frac{1}{4\pi}
		\left[
		\frac{1}{r_h}-8\pi r_h\rho(r_h)
		\right].
		\label{eq:main_temperature}
	\end{equation}
	The first term is the usual geometric contribution evaluated at the shifted horizon. The second term is the local atmospheric correction at the horizon. Thus the horizon shift and the temperature shift are not independent assumptions; both follow from the same conserved Hawking-scale source.
	\begin{figure*}[htp]
		\includegraphics[width=16cm]{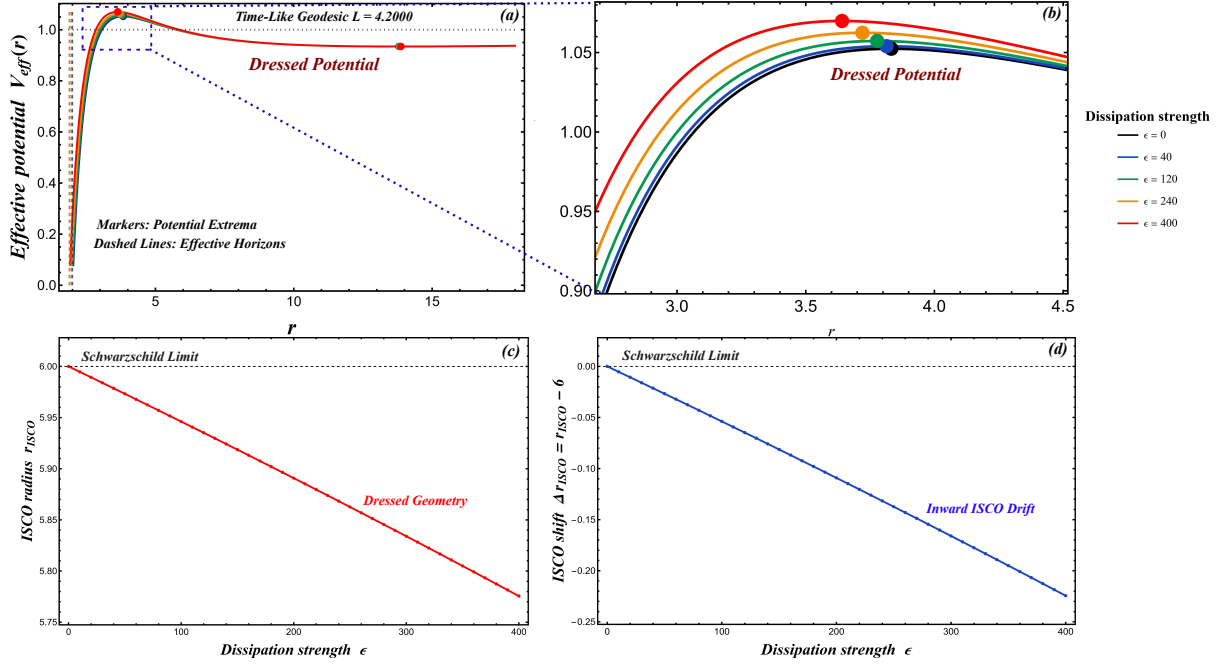}
		\caption{\textbf{Orbital restructuring in the dressed mean geometry.}
			\textbf{(a)} Effective potential \(V_{\mathrm{geo}}(r)=f(r)(1+L^2/r^2)\) for a timelike geodesic with \(L=4.2\). Dashed vertical lines denote the corresponding horizon locations, and markers identify extrema of the potential.
			\textbf{(b)} Enlarged view of the potential barrier for \(\varepsilon=0,\,40,\,120,\,240,\) and \(400\). Increasing dressing strength changes both the height and the position of the barrier.
			\textbf{(c)} ISCO radius \(r_{\mathrm{ISCO}}\) as a function of \(\varepsilon\), obtained from the marginal-stability condition for circular timelike orbits.
			\textbf{(d)} ISCO displacement \(\Delta r_{\mathrm{ISCO}}=r_{\mathrm{ISCO}}-6M\). The inward drift shows that the same lapse renormalization that shifts the horizon also changes the conservative orbital structure.}
		\label{fig3}
	\end{figure*}

	The conservative orbital sector is then determined by geodesic motion in the dressed background. For equatorial timelike motion,
	\begin{equation}
		E=f(r)\dot t,
		\qquad
		L=r^2\dot\phi,
		\qquad
		\dot r^{\,2}=E^2-f(r)\left(1+\frac{L^2}{r^2}\right).
		\label{eq:main_geodesic}
	\end{equation}
	Equivalently, \(\dot r^{\,2}+V_{\rm geo}(r)=E^2\), with
	\begin{equation}
		V_{\rm geo}(r)=f(r)\left(1+\frac{L^2}{r^2}\right).
		\label{eq:main_potential}
	\end{equation}
	The orbital modifications are therefore not introduced by adding a dissipative force to the particle equation of motion. They arise because the lapse itself has changed. The same function \(f(r)\) determines the horizon, the temperature, the height and location of the effective potential barrier, the circular-orbit sequence, and the innermost stable circular orbit.
	
	The stochastic part of the open-system construction is restored only after the mean geometry has been fixed. Writing
	\begin{equation}
		g_{\mu\nu}=\bar g_{\mu\nu}+h_{\mu\nu},
		\label{eq:main_metric_split}
	\end{equation}
	the next-order dynamics has the Einstein--Langevin form
	\begin{equation}
		\delta G_{\mu\nu}[h;\bar g]=8\pi\Xi_{\mu\nu},
		\qquad
		\left\langle
		\Xi_{\mu\nu}(x)\Xi_{\alpha\beta}(x')
		\right\rangle
		=
		N_{\mu\nu\alpha\beta}(x,x').
		\label{eq:main_langevin}
	\end{equation}
	Here \(N_{\mu\nu\alpha\beta}\) is the noise kernel of the coarse-grained environment. The present work does not compute this kernel explicitly. It focuses on the deterministic conserved background, while Eq.~\eqref{eq:main_langevin} identifies the next level at which the dressed black hole becomes a fluctuating spacetime. The detailed algebra leading to the reduced equations, the conserved pressure balance, the ADM normalization, and the formal Einstein--Langevin extension is collected in the appendices.

\section{Mean-field dressing of the Schwarzschild reference geometry}
\begin{figure*}[htp]
	\includegraphics[width=16cm]{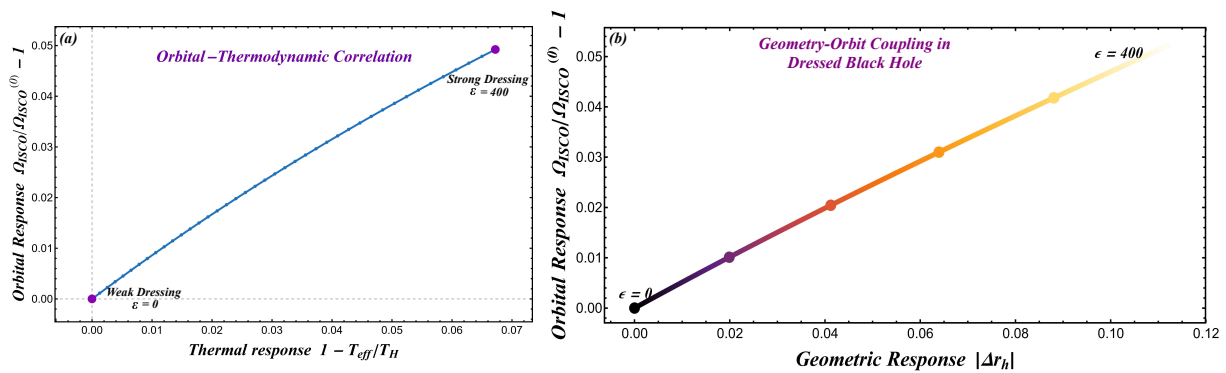}
	\caption{\textbf{Correlated thermal, geometric and orbital response.}
		\textbf{(a)} Normalized orbital response \(\mathcal R_{\rm orb}=|r_{\rm ISCO}(\varepsilon)-6M|/(6M)\) plotted against the thermal deviation \(1-T_{\mathrm{eff}}/T_H\). The monotonic trend indicates that the change in orbital structure tracks the suppression of the effective temperature.
		\textbf{(b)} Normalized orbital response plotted against the magnitude of the horizon shift \(|\Delta r_h|\). The near-linear behavior over the displayed range shows that the orbital response is controlled by the same geometric deformation that moves the horizon.}
	\label{fig4}
\end{figure*}
The first result is the deformation of the Schwarzschild reference geometry by the conserved Hawking-scale atmosphere. The calculation is performed at fixed ADM mass, so the mass measured by an observer at infinity is not changed. What changes is the radial distribution of the gravitating contribution. The effective atmosphere carries part of the fixed mass outside a given radius, and the enclosed mass near the horizon is therefore reduced relative to the Schwarzschild reference value. This redistribution is the reason why the zero of the dressed lapse is shifted inward.

Figure~\ref{fig1} gives a compact visual summary of this mean-field dressing. The embedding comparison in Fig.~\ref{fig1}(a) shows that the dressed surface remains close to the Schwarzschild reference, but the near-horizon region is displaced. This is important for the interpretation of the solution. The construction does not produce a disconnected or violently modified geometry. It gives a controlled exterior deformation generated by a conserved effective source. The dressed geometry therefore remains within the same black-hole class, while acquiring a small but systematic horizon-scale correction.

The causal sketches in Figs.~\ref{fig1}(b) and \ref{fig1}(c) should be read in the same spirit. They are not meant to represent the full stochastic causal structure of an Einstein--Langevin spacetime. Rather, they show how the deterministic horizon scale is displaced before the stochastic sector is restored. The Schwarzschild radius remains the reference surface, while the dressed horizon is the root of the corrected lapse. Thus the horizon shift is not imposed as a boundary condition. It is an output of the mean-field geometry.

The quantitative panel, Fig.~\ref{fig1}(d), confirms this interpretation. The dressed lapse crosses zero at a smaller radius than the Schwarzschild reference, and the same solution gives a reduced effective temperature. These two numbers are physically tied. The horizon position and the surface gravity are obtained from the same lapse. The figure therefore shows a correlated geometric and thermal response: once the atmosphere is included in the mass profile, the horizon and the temperature change together.

This point is useful for avoiding a possible misreading of the result. The inward motion of the horizon does not mean that the black hole has lost ADM mass. The ADM mass is fixed throughout. The shift instead reflects how the effective atmospheric contribution is distributed outside the near-horizon region. In this sense the dressed horizon is smaller because less of the fixed total mass is enclosed at that radius. The effect is therefore a genuine consequence of the chosen ADM normalization, not an artifact of coordinates.

\section{Horizon shift and thermal response}

Figure~\ref{fig2} resolves how the geometry changes as the dressing strength is increased. In the Schwarzschild limit the effective source vanishes, and the reference horizon and Hawking temperature are recovered. Moving away from this limit, the numerical root of the corrected lapse shifts monotonically inward. At weak dressing the numerical curve follows the perturbative result already derived in the theory section and appendix. This agreement is a useful check: it shows that the numerical root is on the expected branch and that the implementation reproduces the analytical small-dressing limit.

The departure from the perturbative line at larger dressing should not be regarded as a failure of the model. It marks the onset of nonlinear response in the root equation. The horizon is determined by the zero of a lapse that contains the atmosphere-dependent radial term. Once the root moves appreciably, the value of the atmospheric contribution at the new root also changes. The response is then no longer captured by the first-order shift alone. Figure~\ref{fig2}(a) therefore separates the perturbative regime from the stronger, but still smooth, mean-field response.

Figure~\ref{fig2}(b) shows the corresponding temperature deviation. The trend is monotonic: stronger dressing suppresses the effective temperature. This is consistent with the geometric result. The surface gravity is controlled by the slope of the corrected lapse at the dressed horizon, and the local effective density contributes directly to that slope. Therefore the suppression is not added by hand as an external thermodynamic assumption. It follows from the same geometry that produced the horizon shift.

The two panels together provide the main thermodynamic message of the paper. The effective atmosphere does not merely move the horizon; it changes the local gravitational field at the horizon. The temperature response is therefore a local diagnostic of the same mean-field dressing. This gives the model a useful internal consistency: geometry and thermodynamics are not fitted independently, but are computed from the same dressed solution.

The label on the horizontal axis of Fig.~\ref{fig2} uses the language of dissipation strength. In the discussion we refer to the same parameter as the dressing strength, because the present results use the deterministic mean field. A true dissipative interpretation would require the explicit nonlocal dissipation kernel of the full Einstein--Langevin problem. That kernel is not computed here, so the safer reading of the plotted parameter is that it measures the strength of the conserved mean atmospheric contribution.

\section{Orbital restructuring in the dressed mean geometry}

The next consequence is orbital. The relevant point is not that particles experience a new force. They do not. The trajectories and circular orbits are geodesics of the dressed mean metric. The atmosphere has already modified the background geometry, and the particle responds to that geometry through the usual conservative geodesic equations. This is why the orbital effects should be described as metric-induced restructuring, not as friction, drag, or stochastic forcing.

Figure~\ref{fig3}(a) shows that the effective potential keeps the familiar black-hole barrier form. This is a necessary consistency check. The dressing does not remove the standard exterior orbital structure. Instead, it changes the scale and location of the potential features. The dashed horizon markers and the extrema of the potential move together as the dressing strength is varied. This behavior is expected because the same lapse controls both the capture boundary and the radial potential.

The enlarged view in Fig.~\ref{fig3}(b) makes the effect more visible. The potential barrier is not simply shifted rigidly. Its height and radial position are both renormalized. This indicates that the atmosphere modifies the shape of the near-horizon exterior geometry, rather than producing only a constant radial offset. The result is still conservative, but the conservative potential itself is changed.

Figures~\ref{fig3}(c) and \ref{fig3}(d) give a sharper diagnostic through the ISCO. The ISCO is a useful quantity because it is determined by the marginal stability of circular timelike orbits and does not depend on the arbitrary starting point of a plunge trajectory. The dressed solution moves the ISCO inward over the displayed parameter range. This inward drift is consistent with the horizon result: the exterior geometry becomes effectively more compact, and the last stable circular orbit follows the same radial reorganization.

The important point is that the ISCO shift is not an independent assumption. It is a consequence of the same lapse renormalization that moved the horizon and changed the temperature. Thus Fig.~\ref{fig3} extends the result from horizon thermodynamics to conservative orbital mechanics. The dressed black hole is not only thermally modified; its exterior geodesic structure is modified in a correlated way.
\begin{figure*}[htp]
	\includegraphics[width=15cm]{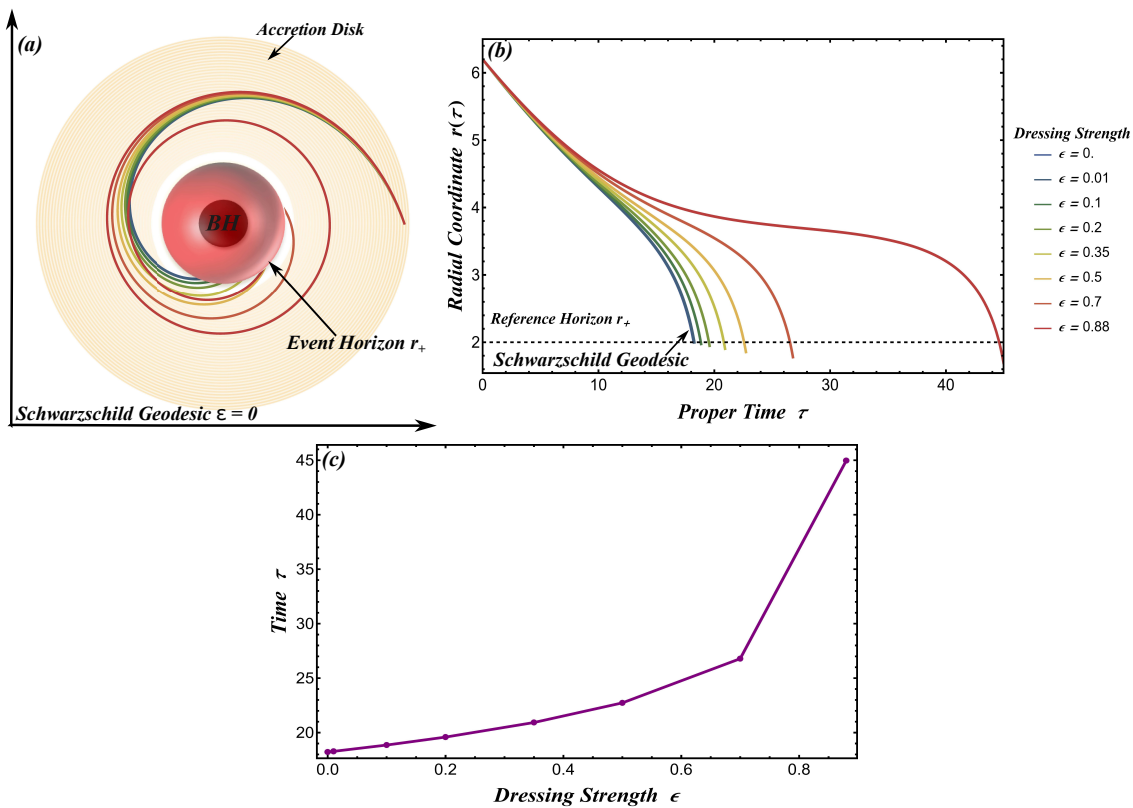}
	\caption{\textbf{Plunge dynamics in the dressed mean geometry.}
		\textbf{(a)} Equatorial plunge trajectories for increasing dressing strength, shown relative to the accretion-disk background and the horizon scale. The Schwarzschild trajectory, \(\varepsilon=0\), provides the reference case.
		\textbf{(b)} Radial evolution \(r(\tau)\) for the same family of trajectories. The dashed horizontal line marks the Schwarzschild reference radius \(r_0=2M\). Larger dressing produces a longer residence time in the near-horizon region before capture.
		\textbf{(c)} Capture time as a function of dressing strength. Capture is defined by the first crossing of a small cutoff outside the dressed horizon. The increase of the capture time reflects geodesic motion in the renormalized lapse, not an additional dissipative force on the particle.}
	\label{fig5}
\end{figure*}
\section{Correlation between thermal, geometric and orbital responses}

Figure~\ref{fig4} tests whether the orbital response is connected to the geometric and thermal responses. The plotted orbital diagnostic is the normalized ISCO displacement defined in the caption and in the surrounding text. This choice is appropriate because the ISCO is a clean circular-orbit observable. It is less dependent on numerical choices than a plunge time and is directly tied to the conservative structure of the metric.

In Fig.~\ref{fig4}(a), the orbital response grows with the thermal deviation. The trend shows that the suppression of the effective temperature and the inward drift of the ISCO are controlled by the same mean-field dressing. They probe different aspects of the dressed lapse: the temperature is sensitive to the near-horizon slope, while the ISCO depends on the radial structure of the potential farther outside. Their correlation therefore gives a nontrivial check that the model produces a coherent geometric response rather than isolated numerical effects.

Figure~\ref{fig4}(b) gives the same message in terms of the horizon displacement. As the magnitude of the horizon shift increases, the orbital response also increases. The trend is close to linear over the displayed range, but it should not be promoted to a universal law. It is a property of the chosen parameter window and the subcritical branch of the solution. At stronger dressing, the nonlinear horizon-root structure would become more important and the relation need not remain linear.

The value of Fig.~\ref{fig4} is therefore conceptual as much as numerical. It shows that the model ties together three sectors that are often discussed separately: the horizon radius, the Hawking temperature, and the timelike orbital structure. Once the atmosphere is encoded in the mass profile, these quantities become different projections of one dressed geometry. This is the central physical message of the results.

\section{Plunge dynamics and near-horizon residence}

The final diagnostic is the plunge dynamics shown in Fig.~\ref{fig5}. This part of the analysis uses the same conservative geodesic prescription as the potential calculation. No stochastic force is added, and no explicit dissipative term is inserted into the radial motion. The purpose of the figure is narrower and cleaner: to show how the deterministic dressed geometry changes the route to capture.

Figure~\ref{fig5}(a) shows equatorial plunge trajectories for increasing dressing strength. The trajectories retain the expected spiral-plunge character, so the black hole remains absorptive. The difference lies in the near-hole segment of the motion. As the dressing grows, the trajectory spends a longer part of its evolution in the near-horizon exterior before crossing the capture cutoff. This is the dynamical counterpart of the potential restructuring in Fig.~\ref{fig3}.

The radial evolution in Fig.~\ref{fig5}(b) shows the same effect more directly. Compared with the Schwarzschild case, the dressed trajectories approach the near-horizon region more slowly in proper time. The dashed reference line marks the Schwarzschild radius, while the actual capture condition is set outside the dressed horizon. Since the dressed horizon is displaced, the particle probes a modified lapse profile before the integration stops. The delay is therefore not a sign of mechanical friction. It reflects geodesic motion in a renormalized near-horizon geometry.

Figure~\ref{fig5}(c) condenses this behavior into the capture time. The monotonic increase of the capture time shows that the dressed mean geometry supports a longer near-horizon residence before absorption. This result is consistent with the previous figures. The same dressing that shifts the horizon and suppresses the temperature also changes the effective potential and the radial infall rate.

Taken together, Figs.~\ref{fig1}--\ref{fig5} establish a single sequence of effects. The conserved Hawking-scale atmosphere modifies the mass profile. The modified mass profile renormalizes the lapse. The renormalized lapse determines the horizon, the surface gravity, the orbital potential, the ISCO, and the plunge time. These are linked consequences of one deterministic mean-field solution. The stochastic Einstein--Langevin sector remains the next level of the theory, where the dressed background itself would acquire noise-induced metric fluctuations.

\section{Conclusion}

We have introduced a Hawking-scale mean-field construction in which the
open-system character of a black hole is encoded at the level of the
gravitational background. The central novelty is to use the thermal scale
set by Hawking radiation as the organizing scale for a conserved effective
atmosphere surrounding the horizon, rather than treating horizon openness
as a property of radiation or probe dynamics alone. In this way, the
quantum-Langevin viewpoint is implemented in a semiclassical gravitational
setting: the coarse-grained environment is not treated as an external noise
or friction force acting on a probe in a fixed spacetime, but first appears
as a conserved source in the Einstein equations and renormalizes the metric
itself. At fixed ADM mass, this source redistributes part of the gravitating
contribution outside the near-horizon region, producing a radially dependent
mass function, a shifted zero of the lapse, and a modified surface gravity.
The horizon displacement and the effective temperature correction are
therefore not separate assumptions, but two consequences of the same
conserved Hawking-scale dressing.

The same corrected lapse also controls the conservative orbital sector.
Timelike geodesics respond to the dressed geometry through the effective
potential $V_{\mathrm{geo}}(r) = f(r)\left(1 + \frac{L^{2}}{r^{2}}\right)$.
As a result, the horizon shift, thermal response, potential barrier, ISCO
drift, and plunge time are linked by the single function $f(r)$. The orbital
changes are not produced by adding an external dissipative force to the
particle equation of motion. They are consequences of geodesic motion in a
background whose near-horizon structure has already been modified by the
coarse-grained quantum atmosphere. This correlation between thermodynamic,
geometric, and orbital responses is the main physical message of the paper:
Hawking-scale openness leaves a coherent, correlated imprint across sectors
that are usually treated separately, providing a direct link between horizon
thermodynamics and observable features of black-hole dynamics.

The construction should be viewed as an intermediate semiclassical regime
between fixed-background treatments and a fully fluctuating
Einstein--Langevin spacetime. In standard semiclassical gravity, one studies
quantum fields on a classical background or the mean backreaction of quantum
expectation values. In the full Einstein--Langevin picture, stress-tensor
fluctuations generate stochastic metric fluctuations through a noise kernel.
The present work isolates the deterministic part of that broader program and
establishes it as the necessary first step: a conserved Hawking-scale source
first defines the dressed mean geometry, while the fluctuating sector can
subsequently be restored by writing$g_{\mu\nu} = \bar{g}_{\mu\nu} + h_{\mu\nu},$
and evolving $h_{\mu\nu}$ with the corresponding stochastic equation. Thus
the novelty is that quantum openness is promoted from a property of radiation
or probe dynamics to a property of the black-hole solution itself, while the
microscopic noise kernel, dissipation kernel and fluctuation--dissipation
relation define the next stage of the theory. This positions the dressed
mean geometry as the controlled starting point for future stochastic gravity
programs addressing horizon-scale openness.

\appendix

\section{Open-system mean field and conserved effective source}

We use units $c=\hbar=k_B=1$. When dimensional clarity is useful, we keep $\ell_{\rm P}^{2}=G$ explicit. The stress tensor appearing in the geometrized Einstein equations is then the physical stress tensor multiplied by $G$. In fully geometrized units one may set $G=\ell_{\rm P}^{2}=1$.

The open-system picture separates the geometry into a deterministic mean sector and a fluctuating sector. Formally one may write
\begin{equation}
	G_{\mu\nu}[g]
	=
	8\pi\left(T^{(m)}_{\mu\nu}+T^{({\rm env})}_{\mu\nu}+\Xi_{\mu\nu}\right),
	\label{app:open_full}
\end{equation}
where $T^{(m)}_{\mu\nu}$ denotes ordinary matter, $T^{({\rm env})}_{\mu\nu}$ is the mean environmental contribution associated with the coarse-grained horizon atmosphere, and $\Xi_{\mu\nu}$ is the stochastic source. The stochastic source has vanishing mean,
\begin{equation}
	\left\langle \Xi_{\mu\nu}(x)\right\rangle=0,
	\label{app:noise_mean}
\end{equation}
but non-zero correlations,
\begin{equation}
	\left\langle \Xi_{\mu\nu}(x)\Xi_{\alpha\beta}(x')\right\rangle
	=
	N_{\mu\nu\alpha\beta}(x,x'),
	\label{app:noise_kernel}
\end{equation}
where $N_{\mu\nu\alpha\beta}$ is the noise kernel. The present work uses the deterministic part of this structure. Defining
\begin{equation}
	\bar g_{\mu\nu}\equiv \left\langle g_{\mu\nu}\right\rangle,
	\label{app:mean_metric_def}
\end{equation}
and taking the exterior ordinary-matter sector to vanish, Eq.~\eqref{app:open_full} reduces at the mean-field level to
\begin{equation}
	G_{\mu\nu}[\bar g]
	=
	8\pi T^{({\rm env})}_{\mu\nu}.
	\label{app:mean_einstein}
\end{equation}
The Bianchi identity imposes
\begin{equation}
	\bar\nabla_\mu T^{\mu\nu}_{({\rm env})}=0.
	\label{app:mean_conservation}
\end{equation}
This is the main consistency condition. Once the horizon atmosphere is placed on the right-hand side of the Einstein equation, it cannot be chosen as an arbitrary radial profile. It must define a conserved effective source. The stochastic sector is discussed again in Appendix~\ref{app:EL_extension}; all results in the main text refer to the deterministic mean geometry.

\section{Static spherical reduction}

The static, spherically symmetric metric is written as
\begin{equation}
	ds^2=-e^{2\Phi(r)}f(r)\,dt^2
	+
	\frac{dr^2}{f(r)}
	+
	r^2d\Omega^2,
	\qquad
	f(r)=1-\frac{2m(r)}{r}.
	\label{met:metric}
\end{equation}
The function $m(r)$ is the mass function, while $\Phi(r)$ describes any redshift not already contained in the lapse $f(r)$. The Schwarzschild limit is
\begin{equation}
	m(r)=M,
	\qquad
	\Phi(r)=0,
	\qquad
	f(r)=1-\frac{2M}{r}.
	\label{app:schwarzschild_limit}
\end{equation}
The source is written as an anisotropic effective fluid,
\begin{equation}
	T^\mu{}_{\nu}
	=
	\mathrm{diag}\left[-\rho(r),p_r(r),p_\perp(r),p_\perp(r)\right].
	\label{met:stress}
\end{equation}
For the metric in Eq.~\eqref{met:metric}, the independent mixed Einstein-tensor components are
\begin{equation}
	G^t{}_{t}
	=
	-\frac{1-f-rf'}{r^2},
	\label{app:Gtt}
\end{equation}
\begin{equation}
	G^r{}_{r}
	=
	-\frac{1-f}{r^2}
	+
	\frac{f'}{r}
	+
	\frac{2f\Phi'}{r},
	\label{app:Grr}
\end{equation}
and
\begin{equation}
	G^\theta{}_{\theta}
	=
	G^\phi{}_{\phi}
	=
	f\left(\Phi''+\Phi'^2+\frac{\Phi'}{r}\right)
	+
	\frac{3}{2}f'\Phi'
	+
	\frac{f'}{r}
	+
	\frac{f''}{2}.
	\label{app:Gthetatheta}
\end{equation}
The term $f'/r$ in Eq.~\eqref{app:Grr} is important. If it is omitted, the intermediate redshift equation is wrong.

Using
\begin{equation}
	f(r)=1-\frac{2m(r)}{r},
	\qquad
	f'(r)=-\frac{2m'(r)}{r}+\frac{2m(r)}{r^2},
	\label{app:fprime_m}
\end{equation}
Eq.~\eqref{app:Gtt} gives
\begin{equation}
	G^t{}_{t}=-\frac{2m'(r)}{r^2}.
	\label{app:Gtt_mass}
\end{equation}
Since $T^t{}_{t}=-\rho(r)$, the $tt$ equation gives
\begin{equation}
	m'(r)=4\pi r^2\rho(r).
	\label{met:mprime}
\end{equation}
This equation defines $m(r)$ as the gravitating mass enclosed inside radius $r$.

The radial equation follows from $G^r{}_{r}=8\pi p_r$. Substituting Eq.~\eqref{app:fprime_m} into Eq.~\eqref{app:Grr} gives
\begin{equation}
	-\frac{2m'(r)}{r^2}
	+
	\frac{2f(r)\Phi'(r)}{r}
	=
	8\pi p_r(r).
	\label{app:rr_intermediate}
\end{equation}
Using Eq.~\eqref{met:mprime}, one obtains
\begin{equation}
	\Phi'(r)
	=
	\frac{4\pi r[\rho(r)+p_r(r)]}{f(r)}.
	\label{met:phiprime}
\end{equation}
Thus $\rho(r)$ changes the mass profile, whereas the combination $\rho+p_r$ decides whether the dressing also produces an additional temporal redshift.

\section{Conservation equation and anisotropic balance}

The conservation equation follows from the Bianchi identity,
\begin{equation}
	\nabla_\mu T^\mu{}_{\nu}=0.
	\label{met:conservation}
\end{equation}
Only the radial component gives a nontrivial condition. Direct evaluation gives
\begin{equation}
	p_r'(r)
	=
	-[\rho(r)+p_r(r)]
	\left[
	\Phi'(r)+\frac{f'(r)}{2f(r)}
	\right]
	+
	\frac{2}{r}\left[p_\perp(r)-p_r(r)\right].
	\label{met:tov}
\end{equation}
This is the anisotropic hydrostatic-balance equation for the effective atmosphere. It fixes the allowed pressure anisotropy once the density profile and radial pressure are specified. It is often useful to write the same equation in a mass-function form. Combining Eqs.~\eqref{app:fprime_m}, \eqref{met:mprime}, and \eqref{met:phiprime} gives
\begin{equation}
	\Phi'(r)+\frac{f'(r)}{2f(r)}
	=
	\frac{m(r)+4\pi r^3p_r(r)}{r^2f(r)}.
	\label{app:tov_identity}
\end{equation}
Equation~\eqref{met:tov} is therefore equivalent to
\begin{equation}
	p_r'(r)
	=
	-[\rho(r)+p_r(r)]
	\frac{m(r)+4\pi r^3p_r(r)}{r^2f(r)}
	+
	\frac{2}{r}\left[p_\perp(r)-p_r(r)\right].
	\label{app:tov_mass_form}
\end{equation}
The construction used in the main text is the special conserved branch of this general anisotropic system.

\section{Hawking-scale source and pressure balance}

The Schwarzschild reference geometry provides the two scales
\begin{equation}
	r_0=2M,
	\qquad
	T_H=\frac{1}{8\pi M}.
	\label{met:scales}
\end{equation}
Here $M$ is the ADM mass measured at spatial infinity. The Hawking temperature provides the natural scale for the coarse-grained atmosphere. We take the geometrized effective density to be
\begin{equation}
	\rho(r)
	=
	\varepsilon\,\ell_{\rm P}^{2}T_H^4
	\left(\frac{r_0}{r}\right)^n,
	\qquad
	n>3.
	\label{met:rho}
\end{equation}
The parameter $\varepsilon$ is dimensionless and measures the strength of the coarse-grained environmental contribution. It may absorb numerical constants, greybody factors and the effective number of near-horizon degrees of freedom. The restriction $n>3$ ensures that the integrated atmospheric contribution is finite at spatial infinity. The profile should be understood as an exterior effective atmosphere. The present calculation does not use it as a regular interior solution all the way to $r=0$.

For the lapse-only branch used in the main text we take
\begin{equation}
	p_r(r)=-\rho(r).
	\label{met:pr}
\end{equation}
Then $\rho+p_r=0$, so Eq.~\eqref{met:phiprime} gives
\begin{equation}
	\Phi'(r)=0.
	\label{met:phi_zero_derivative}
\end{equation}
After fixing the normalization of the time coordinate at infinity, one sets
\begin{equation}
	\Phi(r)=0.
	\label{app:phi_zero}
\end{equation}
The transverse pressure is not free. It is fixed by conservation. Since
\begin{equation}
	\rho'(r)=-\frac{n}{r}\rho(r),
	\qquad
	p_r'(r)=\frac{n}{r}\rho(r),
	\label{met:rho_derivative}
\end{equation}
and the term proportional to $\rho+p_r$ vanishes in Eq.~\eqref{met:tov}, one obtains
\begin{equation}
	\frac{n}{r}\rho(r)
	=
	\frac{2}{r}\left[p_\perp(r)+\rho(r)\right].
	\label{met:pperp_step}
\end{equation}
Therefore
\begin{equation}
	p_\perp(r)
	=
	\left(\frac{n}{2}-1\right)\rho(r).
	\label{met:pperp}
\end{equation}
The conserved effective stress tensor is then
\begin{equation}
	T^\mu{}_{\nu}
	=
	\mathrm{diag}\left[-\rho,-\rho,
	\left(\frac{n}{2}-1\right)\rho,
	\left(\frac{n}{2}-1\right)\rho\right].
	\label{app:conserved_stress_final}
\end{equation}
This is the key consistency condition. Once the density profile and the branch $p_r=-\rho$ are specified, the transverse pressure is fixed by the Bianchi identity.

\section{ADM normalization of the mass function}

The mass parameter $M$ is defined as the ADM mass. Hence
\begin{equation}
	m(\infty)=M.
	\label{met:adm}
\end{equation}
Using Eq.~\eqref{met:mprime}, the mass enclosed inside radius $r$ is
\begin{equation}
	m(r)
	=
	M-
	\int_r^\infty 4\pi \tilde r^{2}\rho(\tilde r)\,d\tilde r.
	\label{met:mass_integral}
\end{equation}
Substitution of Eq.~\eqref{met:rho} gives
\begin{equation}
	m(r)
	=
	M
	-
	4\pi\varepsilon\ell_{\rm P}^{2}T_H^4 r_0^n
	\int_r^\infty \tilde r^{2-n}\,d\tilde r.
	\label{app:mass_integral_step}
\end{equation}
For $n>3$,
\begin{equation}
	\int_r^\infty \tilde r^{2-n}\,d\tilde r
	=
	\frac{r^{3-n}}{n-3},
	\label{app:power_integral}
\end{equation}
and therefore
\begin{equation}
	m(r)
	=
	M
	-
	\frac{4\pi\varepsilon\ell_{\rm P}^{2}T_H^4 r_0^n}{n-3}
	r^{3-n}.
	\label{met:mass}
\end{equation}
The lapse becomes
\begin{equation}
	f(r)
	=
	1-
	\frac{2M}{r}
	+
	\frac{8\pi\varepsilon\ell_{\rm P}^{2}T_H^4 r_0^n}{n-3}
	r^{2-n}.
	\label{met:lapse}
\end{equation}
This is the corrected mean-field lapse used throughout the paper. The ADM normalization is important. If the mass function is instead anchored at $r_0=2M$, then $f(r_0)=0$ remains true by construction and the horizon displacement is hidden. With ADM normalization, the total mass measured at infinity is fixed. The effective atmosphere redistributes part of that mass outside radius $r$, so the enclosed mass near the horizon is smaller than $M$.

It is useful to introduce the abbreviation
\begin{equation}
	A
	\equiv
	\frac{8\pi\varepsilon\ell_{\rm P}^{2}T_H^4 r_0^n}{n-3},
	\label{app:A_def}
\end{equation}
so that
\begin{equation}
	f(r)=1-\frac{2M}{r}+A r^{2-n}.
	\label{app:lapse_A}
\end{equation}
This form is used below when discussing the horizon roots.

\section{Horizon roots, allowed range and perturbative shift}

The dressed horizon is the largest positive root of
\begin{equation}
	f(r_h)=0.
	\label{met:horizon_def}
\end{equation}
For the lapse in Eq.~\eqref{app:lapse_A}, the derivative is
\begin{equation}
	f'(r)=\frac{2M}{r^2}+(2-n)A r^{1-n}.
	\label{app:fprime_A}
\end{equation}
When $A>0$ and $n>3$, the lapse can have two positive roots, a degenerate root, or no positive root, depending on the strength of the dressing. The critical case is obtained by solving
\begin{equation}
	f(r_c)=0,
	\qquad
	f'(r_c)=0.
	\label{app:critical_conditions}
\end{equation}
From Eq.~\eqref{app:fprime_A},
\begin{equation}
	A
	=
	\frac{2M}{n-2}r_c^{n-3}.
	\label{app:A_from_derivative}
\end{equation}
Substitution into $f(r_c)=0$ gives
\begin{equation}
	1-
	\frac{2M}{r_c}
	+
	\frac{2M}{(n-2)r_c}=0,
	\label{app:critical_step}
\end{equation}
so that
\begin{equation}
	r_c=\frac{2M(n-3)}{n-2}.
	\label{app:rc}
\end{equation}
The corresponding critical value is
\begin{equation}
	A_{\rm crit}
	=
	\frac{2M}{n-2}
	\left[\frac{2M(n-3)}{n-2}\right]^{n-3}.
	\label{app:Acrit}
\end{equation}
Thus $0<A<A_{\rm crit}$ gives two positive roots, and the outer root is the horizon used in the main text. At $A=A_{\rm crit}$ the roots merge and the surface gravity vanishes. For $A>A_{\rm crit}$ the exterior horizon is lost. The parameter range used in the figures lies on the subcritical branch where the outer dressed horizon is well defined.

For weak dressing, write
\begin{equation}
	r_h=r_0+\delta r_h,
	\qquad
	|\delta r_h|\ll r_0.
	\label{met:horizon_expansion}
\end{equation}
Separate the lapse into
\begin{equation}
	f(r)=f_0(r)+\delta f(r),
	\qquad
	f_0(r)=1-\frac{r_0}{r},
	\label{app:lapse_split1}
\end{equation}
with
\begin{equation}
	\delta f(r)
	=
	\frac{8\pi\varepsilon\ell_{\rm P}^{2}T_H^4 r_0^n}{n-3}
	r^{2-n}.
	\label{met:lapse_split}
\end{equation}
Expanding $f(r_h)=0$ around $r_0$ gives
\begin{equation}
	0
	=
	f_0(r_0)
	+
	\delta r_h f_0'(r_0)
	+
	\delta f(r_0)
	+
	\mathcal{O}(\varepsilon^2).
	\label{met:horizon_expand}
\end{equation}
Since $f_0(r_0)=0$ and $f_0'(r_0)=1/r_0$, one obtains
\begin{equation}
	\delta r_h
	=
	-r_0\delta f(r_0)
	=
	-\frac{8\pi\varepsilon\ell_{\rm P}^{2}T_H^4 r_0^3}{n-3}
	+
	\mathcal{O}(\varepsilon^2).
	\label{met:horizon_shift}
\end{equation}
The shift is inward for positive $\varepsilon$ at fixed ADM mass. This does not mean that the total mass has decreased. It means that part of the fixed ADM mass is distributed in the effective atmosphere outside the near-horizon region, so the radius at which $2m(r)=r$ is reduced.

\section{Effective temperature}

For the lapse-only metric,
\begin{equation}
	ds^2=-f(r)dt^2+\frac{dr^2}{f(r)}+r^2d\Omega^2,
	\label{met:lapse_metric}
\end{equation}
the surface gravity is
\begin{equation}
	\kappa=\frac{1}{2}f'(r_h).
	\label{met:kappa}
\end{equation}
The corresponding temperature is
\begin{equation}
	T_{\rm eff}=\frac{\kappa}{2\pi}=\frac{f'(r_h)}{4\pi}.
	\label{met:temperature_def}
\end{equation}
To express this in terms of the local density, differentiate
\begin{equation}
	f(r)=1-\frac{2m(r)}{r}.
	\label{met:f_m}
\end{equation}
This gives
\begin{equation}
	f'(r)
	=
	-\frac{2m'(r)}{r}
	+
	\frac{2m(r)}{r^2}.
	\label{met:fprime}
\end{equation}
At the horizon $f(r_h)=0$, so $m(r_h)=r_h/2$. Therefore
\begin{equation}
	f'(r_h)
	=
	\frac{1}{r_h}
	-
	\frac{2m'(r_h)}{r_h}.
	\label{met:fprime_horizon}
\end{equation}
Using $m'(r_h)=4\pi r_h^2\rho(r_h)$, one obtains
\begin{equation}
	T_{\rm eff}
	=
	\frac{1}{4\pi}
	\left[
	\frac{1}{r_h}
	-
	8\pi r_h\rho(r_h)
	\right].
	\label{met:temperature}
\end{equation}
This formula shows explicitly that the effective temperature is changed by both the displacement of the horizon and the local Hawking-scale energy density at that horizon.

For the more general metric in Eq.~\eqref{met:metric}, with nonzero $\Phi(r)$, the surface gravity is
\begin{equation}
	\kappa
	=
	\frac{1}{2}e^{\Phi(r_h)}f'(r_h),
	\label{app:kappa_general}
\end{equation}
and the temperature is
\begin{equation}
	T_{\rm eff}
	=
	\frac{e^{\Phi(r_h)}}{4\pi}
	\left[
	\frac{1}{r_h}
	-
	8\pi r_h\rho(r_h)
	\right].
	\label{app:T_general}
\end{equation}
The lapse-only branch used in the numerical results is obtained by setting $\Phi=0$.

\section{Trace, curvature and energy-condition diagnostics}

The trace of the conserved effective source in Eq.~\eqref{app:conserved_stress_final} is
\begin{equation}
	T^\mu{}_{\mu}
	=
	-\rho+p_r+2p_\perp
	=
	(n-4)\rho.
	\label{app:trace}
\end{equation}
The Ricci scalar follows from the trace of Einstein's equation,
\begin{equation}
	R=-8\pi T^\mu{}_{\mu},
	\label{app:R_trace_relation}
\end{equation}
so that
\begin{equation}
	R(r)=-8\pi(n-4)\rho(r).
	\label{app:Ricci_scalar}
\end{equation}
For $n=4$ the effective source is traceless. For the values $n>4$ used in the figures, the Ricci scalar is negative for positive $\rho$.

The null-energy combinations are
\begin{equation}
	\rho+p_r=0,
	\qquad
	\rho+p_\perp=\frac{n}{2}\rho.
	\label{app:NEC}
\end{equation}
Thus the radial null-energy condition is saturated, while the tangential one is satisfied for $\rho>0$ and $n>0$. The dominant energy condition requires $\rho\geq |p_i|$. Since
\begin{equation}
	|p_\perp|=\left(\frac{n}{2}-1\right)\rho,
	\label{app:DEC_step}
\end{equation}
it is violated for $n>4$. This violation is not unexpected for an effective quantum atmosphere; the source is not introduced as an ordinary classical perfect fluid, but as a geometrized coarse-grained contribution motivated by semiclassical horizon physics.

\section{Geodesic motion in the dressed mean geometry}

For equatorial motion, $\theta=\pi/2$, the metric is
\begin{equation}
	ds^2
	=
	-e^{2\Phi(r)}f(r)dt^2
	+
	\frac{dr^2}{f(r)}
	+
	r^2d\phi^2.
	\label{app:equatorial_metric}
\end{equation}
The point-particle Lagrangian is
\begin{equation}
	\mathcal{L}
	=
	\frac{1}{2}
	\left[
	-e^{2\Phi}f\dot t^{2}
	+
	\frac{\dot r^{2}}{f}
	+
	r^2\dot\phi^{2}
	\right].
	\label{app:particle_lagrangian}
\end{equation}
Because the metric is independent of $t$ and $\phi$, the conserved quantities are
\begin{equation}
	E=e^{2\Phi}f\dot t,
	\qquad
	L=r^2\dot\phi.
	\label{met:EL}
\end{equation}
The normalization $u^\mu u_\mu=-1$ gives
\begin{equation}
	\dot r^{2}
	=
	E^2e^{-2\Phi(r)}
	-
	f(r)\left(1+\frac{L^2}{r^2}\right).
	\label{app:radial_general}
\end{equation}
Equivalently,
\begin{equation}
	\dot r^{2}+V_{\rm geo}(r)
	=
	E^2e^{-2\Phi(r)},
	\qquad
	V_{\rm geo}(r)
	=
	f(r)\left(1+\frac{L^2}{r^2}\right).
	\label{app:potential_general}
\end{equation}
For the conserved lapse-only branch, $\Phi=0$, this reduces to
\begin{equation}
	\dot r^{2}
	=
	E^2
	-
	f(r)\left(1+\frac{L^2}{r^2}\right),
	\label{met:radial}
\end{equation}
with
\begin{equation}
	V_{\rm geo}(r)
	=
	f(r)\left(1+\frac{L^2}{r^2}\right).
	\label{met:potential}
\end{equation}
Thus the orbital effects arise from geodesic motion in the dressed metric. No external drag force is added to the particle equation of motion.

The corresponding Hamiltonian is
\begin{equation}
	H_{\rm geo}
	=
	\frac{1}{2}g^{\mu\nu}p_\mu p_\nu
	=
	\frac{1}{2}
	\left[
	-\frac{p_t^2}{e^{2\Phi}f}
	+
	fp_r^2
	+
	\frac{L^2}{r^2}
	\right].
	\label{app:hamiltonian}
\end{equation}
This form is useful for checking that the dynamics remains conservative in the deterministic mean geometry.

\section{Circular orbits and ISCO condition}

For the lapse-only branch, circular timelike orbits obey
\begin{equation}
	\dot r=0,
	\qquad
	\frac{dV_{\rm geo}}{dr}=0.
	\label{app:circular_conditions}
\end{equation}
The first condition gives
\begin{equation}
	E^2
	=
	f(r)\left(1+\frac{L^2}{r^2}\right).
	\label{app:E_circular_step}
\end{equation}
The second condition gives
\begin{equation}
	f'(r)\left(1+\frac{L^2}{r^2}\right)
	-
	\frac{2f(r)L^2}{r^3}=0.
	\label{app:dV_step}
\end{equation}
Solving Eq.~\eqref{app:dV_step} for $L^2$ yields
\begin{equation}
	L^2(r)
	=
	\frac{r^3f'(r)}{2f(r)-rf'(r)}.
	\label{app:L2_circ}
\end{equation}
Substitution into Eq.~\eqref{app:E_circular_step} gives
\begin{equation}
	E^2(r)
	=
	\frac{2f(r)^2}{2f(r)-rf'(r)}.
	\label{app:E2_circ}
\end{equation}
Equations~\eqref{app:L2_circ} and \eqref{app:E2_circ} are the circular-orbit expressions used in the main text.

The innermost stable circular orbit is determined by marginal stability,
\begin{equation}
	\frac{dL^2(r)}{dr}=0.
	\label{met:isco_dL}
\end{equation}
To make the condition explicit, differentiate Eq.~\eqref{app:L2_circ}. Since
\begin{equation}
	L^2(r)=\frac{N(r)}{D(r)},
	\qquad
	N(r)=r^3f'(r),
	\qquad
	D(r)=2f(r)-rf'(r),
	\label{app:ND_def}
\end{equation}
the marginal condition is
\begin{equation}
	N'(r)D(r)-N(r)D'(r)=0.
	\label{app:isco_derivative_step}
\end{equation}
Using
\begin{equation}
	N'(r)=3r^2f'(r)+r^3f''(r),
	\qquad
	D'(r)=f'(r)-rf''(r),
	\label{app:ND_derivatives}
\end{equation}
Eq.~\eqref{app:isco_derivative_step} reduces to
\begin{equation}
	r f(r)f''(r)+3f(r)f'(r)-2r f'(r)^2=0.
	\label{met:isco}
\end{equation}
In the Schwarzschild limit $f(r)=1-2M/r$, Eq.~\eqref{met:isco} gives
\begin{equation}
	r_{\rm ISCO}=6M.
	\label{app:Schw_ISCO}
\end{equation}
For the dressed lapse in Eq.~\eqref{met:lapse}, the same condition determines the corrected ISCO radius. The normalized orbital response used to compare with the thermal and horizon diagnostics may be written as
\begin{equation}
	\mathcal{R}_{\rm orb}
	=
	\frac{|r_{\rm ISCO}(\varepsilon)-6M|}{6M}.
	\label{app:orbital_response}
\end{equation}

\section{Plunge trajectories and capture time}

The plunge trajectories are obtained by integrating the first-order geodesic equations for the lapse-only geometry,
\begin{equation}
	\frac{dr}{d\tau}
	=
	-\sqrt{
		E^2
		-
		f(r)\left(1+\frac{L^2}{r^2}\right)
	},
	\qquad
	\frac{d\phi}{d\tau}=\frac{L}{r^2}.
	\label{met:plunge}
\end{equation}
The negative branch selects inward motion. Capture is defined by the first crossing of
\begin{equation}
	r(\tau)=r_h+\Delta r,
	\label{met:capture}
\end{equation}
where $\Delta r$ is a small numerical cutoff outside the dressed horizon. Equivalently,
\begin{equation}
	\tau_{\rm cap}
	=
	\inf\left\{\tau\,|\,r(\tau)=r_h+\Delta r\right\}.
	\label{app:capture_time_def}
\end{equation}
The same initial radius, conserved energy, angular momentum and cutoff are used when comparing different values of $\varepsilon$. No stochastic or dissipative force is included in these trajectories. The capture-time shift therefore comes from the corrected lapse $f(r)$, not from an added drag term.

\section{Formal Einstein--Langevin extension}
\label{app:EL_extension}

The deterministic metric derived above is the mean field. The stochastic part of the open-system construction is restored by writing
\begin{equation}
	g_{\mu\nu}=\bar g_{\mu\nu}+h_{\mu\nu}.
	\label{app:g_split}
\end{equation}
Linearizing the field equation around the mean background gives the schematic Einstein--Langevin equation
\begin{equation}
	\delta G_{\mu\nu}[h;\bar g]
	=
	8\pi\Xi_{\mu\nu},
	\label{app:EL_schematic}
\end{equation}
with
\begin{equation}
	\left\langle\Xi_{\mu\nu}(x)\Xi_{\alpha\beta}(x')\right\rangle
	=
	N_{\mu\nu\alpha\beta}(x,x').
	\label{app:EL_noise}
\end{equation}
A more complete linear response form would contain a nonlocal dissipation kernel,
\begin{equation}
	\delta G_{\mu\nu}[h;\bar g]
	-
	8\pi\int d^4x'\,
	D_{\mu\nu}{}^{\alpha\beta}(x,x')h_{\alpha\beta}(x')
	=
	8\pi\Xi_{\mu\nu}(x),
	\label{app:EL_dissipation}
\end{equation}
together with the noise correlator in Eq.~\eqref{app:EL_noise}. The present work does not compute $N_{\mu\nu\alpha\beta}$ or $D_{\mu\nu}{}^{\alpha\beta}$ microscopically. Instead, it identifies the conserved deterministic mean geometry on which such stochastic metric fluctuations would propagate. This separation is important: the mean-field solution is deterministic, whereas the full Einstein--Langevin problem requires the noise kernel, the dissipation kernel and their fluctuation--dissipation relation.

\section{Toward slowly rotating effective solutions}

A slowly rotating extension may be constructed using a Hartle-type ansatz,
\begin{equation}
	ds^2
	=
	-e^{2\Phi(r)}f(r)dt^2
	+
	\frac{dr^2}{f(r)}
	+
	r^2\left[d\theta^2+
	\sin^2\theta\left(d\phi-\varpi(r)dt\right)^2\right],
	\label{app:slow_rotation_ansatz}
\end{equation}
where $\varpi(r)$ is the frame-dragging function. In vacuum, $\varpi(r)\sim 2J/r^3$, reproducing the slow-rotation limit of Kerr. In the present framework, the rotational equation would be modified by the rotational sector of the effective atmosphere. This extension is not developed in this paper. The static solution derived above provides the seed geometry for such a perturbative construction.

\section{Summary of the appendix}

The construction proceeds in a fixed order. First, the horizon-scale environment is represented by a conserved effective source in the mean Einstein equation. Second, spherical symmetry reduces the problem to the mass function $m(r)$ and redshift function $\Phi(r)$. Third, the lapse-only branch $p_r=-\rho$ gives $\Phi=0$, while stress-energy conservation fixes $p_\perp=(n/2-1)\rho$. Fourth, the ADM condition $m(\infty)=M$ gives the corrected mass profile and lapse. Fifth, the outer horizon, effective temperature, ISCO and plunge dynamics all follow from the same function $f(r)$. The physical message is that the Hawking-scale open-system contribution is encoded first in the deterministic mean geometry, not as an external force acting on a test particle. The stochastic Einstein--Langevin sector then provides the next stage, where the dressed background itself acquires noise-induced metric fluctuations.

	\section*{Acknowledgments}
	K.A.Y. acknowledges support from the Research Fund for International Young Scientists of the NSFC under Grant No. KYZ04Y22050, Zhejiang Normal University research funding under grant No. ZC304021914 and Zhejiang province postdoctoral research project under grant number ZC304021952.

	\section*{Competing interests}
	The authors declare no competing interests.

	\section*{Data availability}
All data reported in this work are generated from the analytical expressions and solutions presented in the main text and appendices.
	
	\section*{Code availability}
All results illustrated in this work are generated from the analytical expressions given in the article and appendices. The detailed code is available from the corresponding author upon reasonable request.

\end{document}